# Nature of Phase Transitions of Superconducting Wire Networks in a Magnetic Field

X.S. Ling[1], H.J. Lezec[*,2], M.J. Higgins[1], J.S. Tsai[2], J. Fujita[2], H. Numata[2], Y. Nakamura[2], Y. Ochiai[2], Chao Tang[1],
P.M. Chaikin[1,3], and S. Bhattacharya[1]

[1]*NEC Research Institute, 4 Independence Way, Princeton, New Jersey 08540*
[2]*NEC Fundamental Research Laboratories, 34 Myukigaoka, Tsukuba, Japan*
[3]*Department of Physics, Princeton University, Princeton, New Jersey 08544*

(December 4, 1995)

We study I-V characteristics of periodic square Nb wire networks as a function of temperature in a transverse magnetic field, with a focus on three fillings 2/5, 1/2, and 0.618 that represent very different levels of incommensurability. For all three fillings, a scaling behavior of I-V characteristics is found, suggesting a finite temperature continuous superconducting phase transition. The low-temperature I-V characteristics are found to have an exponential form, indicative of the domain-wall excitations.

The presence of a quenched symmetry-breaking field is known to have important consequences on the ordering of low temperature phases in many physical systems. A striking example is the pinning of a two-dimensional (2D) vortex lattice by a periodic potential in a superconducting wire network. Without pinning, a 2D elastic vortex lattice in homogeneous superconducting thin films would not have long-range translational order at any finite temperature [1] and cannot have long-range superconducting phase coherence even at $T = 0$. A periodic pinning potential, however, when it is commensurate to the vortex lattice, can induce a gap in the low-energy excitation spectrum [2] and create a new thermodynamic phase [3] of a pinned 2D solid with true long-range translational order, and with superconducting phase coherence [4]. In the presence of a high-order commensurate (or incommensurate) potential, the competition between the vortex-vortex interactions and the vortex-network interactions leads to a whole new class of problems [5–7]. For example, the vortex lattice may: (a) become a 2D "floating" solid and again lose its translational order and superconducting phase coherence at any temperature [5]; (b) form a metastable "glassy" phase [6]; (c) be pinned in a commensurate phase [7] and thus superconducting at low temperatures. The issue is far from being settled.

Closely related to the low-temperature thermodynamic phase, the nature of the superconducting transition of a superconducting network in a magnetic field is also not well understood. At filling $f = 1/2$, where $f$ is the fraction of a flux quantum $\phi_0 = hc/2e$ per plaquette, for example, the vortex configuration of the ground state of the system has a checker-board pattern [4]. Thus, the ground state of the system has the discrete symmetry of the two-fold degeneracy as well as the continuous symmetry of an arbitrary global phase change. Two types of excitations are possible: vortex-antivortex pairs of the continuous phase variable and domain walls of the two ground states. If the two types of excitations do not interact, they should lead to two independent transitions [9]: a Kosterlitz-Thouless (KT) transition [8] in the underlying network and an Ising melting transition in the vortex lattice. The superconducting transition will be determined by the one with the lower transition-temperature. Interesting physics arises when the two types of excitations do couple [10]. Due to the screening of the vortex interactions by the domain walls, when the domain wall energy goes to zero at an Ising-like transition, a spontaneous generation of domain walls will unbind the vortex pairs and induce a KT-like transition [10]. In this scenario there will be a single superconducting transition of a new universality class [10,11]. In spite of intense theoretical effort on this problem, the exact nature of this transition remains unknown [11]. We show in this paper that *the superconducting transition at $f = 1/2$ is continuous and is strongly influenced by the Ising-like excitations (domain walls), but is not an Ising transition.* Moreover, we find that *the phase transitions at higher order commensurate or incommensurate fillings such as $f = 2/5$ and $f = 0.618$ are, surprisingly, also similar in nature.*

We describe here an experiment on a periodic square superconducting wire network. The network sample in our experiment is fabricated from high-quality sputtered Nb film. The microfabrication consists of standard steps, i.e., electron beam lithography, lift-off and reactive ion etch. The final sample has 800 × 800 cells, with a lattice constant 1 $\mu$m, wire width 0.25 $\mu$m and thickness 0.1 $\mu$m. SEM (scanning electron microscope) examinations of the final sample show that the overall periodicity is maintained very well, as shown in Fig. 1. The zero-field transition has $T_c$ (at $0.5R_n$) $\approx$ 8.920 K and width $((10\% - 90\%)R_n) \approx$ 10 mK. The normal-state resistance $R_n \approx 1.2$ $\Omega$/square. The standard four-probe technique is used for the transport measurements. To ensure current uniformity, we use large contact pads (silver paint), each covering the entire edge of the network. The voltage contacts are lithographically patterned with the network, both are 5 $\mu$m in width and are of the same material Nb. The close-loop temperature regulation is done by a Linear Research LR-130 controller and the voltage is measured by a transformer-coupled PAR-124A lock-in amplifier with a square-wave current at 10.2 Hz. To test possible heating effects on the I-V curves, we use different duty cycles and find no measurable differences. The temperature reading accuracy around 9 K is about 1 mK, although the temperature stability is better than 0.5 mK (determined by using the sample itself as a thermometer).

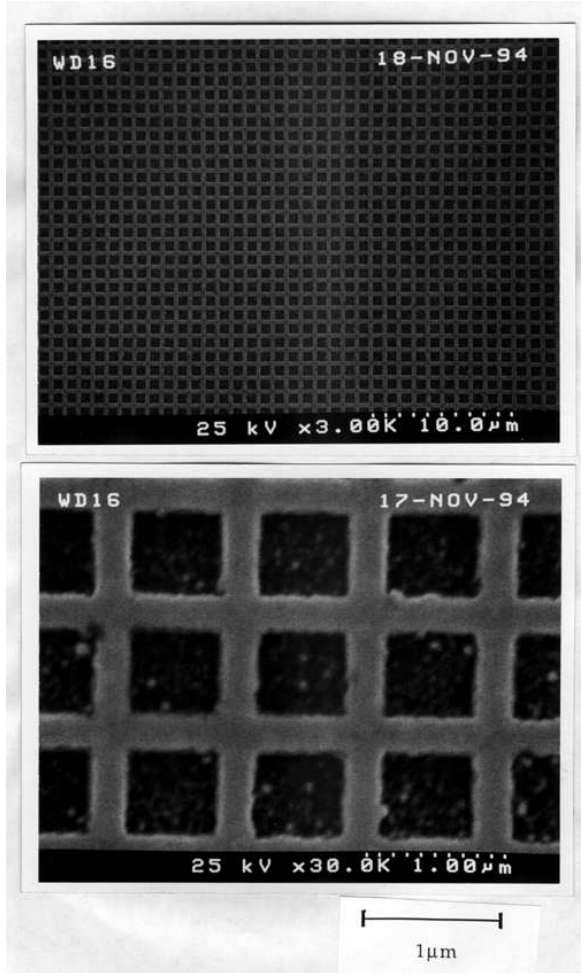

FIG. 1. SEM micrographs showing the structure of the superconducting Nb wire network used in this work.

Fig. 2 shows the sample resistance as a function of magnetic field at various fixed temperatures. As the sample is cooled from the normal state, the sample resistance first dips at $f = 0$ and other integer fillings; upon further cooling, the resistance dips at $f = 1/2$ and other fillings. Taking advantage of the variation of the sample resistance with $f$, one can determine a mean-field phase boundary [12] that resembles some features of the interesting Hofstadter spectrum [13]. That phase boundary is not, however, the true phase boundary of the superconducting transitions due to the neglect of the vortex fluctuations.

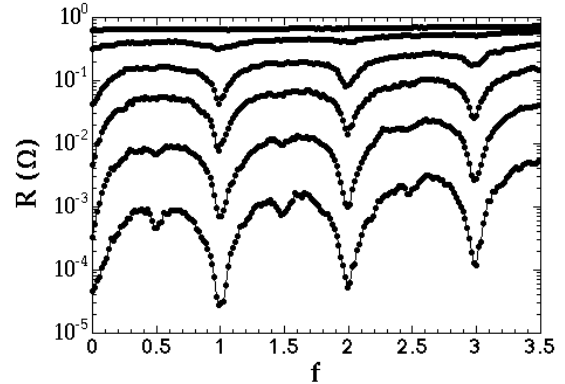

FIG. 2. Sample resistance $R(=V/I)$ as a function of filling factor $f$ at various temperatures: from top, 8.920, 8.912, 8.907, 8.901, 8.895 and 8.890 K. The test current was $I_{ac} = 10\mu A$ at 10.2 Hz.

Since the emergence of long-range superconducting phase coherence is characterized by nonlinear I-V characteristics, we study how the I-V curves behave as the sample is cooled in a fixed field. As a calibration, we measure the temperature dependence of the I-V characteristics at $f = 0, 1,$ and 3 [14]. We find that the I-V curve changes from ohmic (at low current) to power-law with decreasing temperature. The low-temperature power-law I-V curves are characteristic of the vortex-antivortex pairs bound by a logarithmic interaction. The temperature dependence of the power-law exponent $a$ ($V \sim I^a$) shows the familiar jump at $a \approx 3$ [15]. Other aspects of the data for $f = 0, 1,$ and 3 are also consistent with earlier works [15]. We thus conclude that at $f = 0, 1,$ and 3 our system undergoes a KT transition.

In sharp contrast to the power-law I-V curves at $f = 0, 1,$ and 3, the low-temperature I-V characteristics at non-integer fillings are found to have an exponential form and the I-V curves behave similarly for different non-integer fillings as a function of temperature. Fig. 3 shows the I-V curves for $f = 2/5, 1/2, 0.618,$ and $5/2$ [14]. At high temperatures and for all these fillings, the I-V curves are ohmic at low current and concave upward at high current. At low temperatures, the I-V curves become progressively more concave downward. The high-T and low-T behaviors are separated by a $V \sim I^3$ power-law. The lowest temperature curves for all four fillings in Fig. 3 can be fitted with a form $V \sim I \exp(-I_T/I)$, where $I_T$ is a fitting constant. This exponential form can be described by a simple domain-wall nucleation process [16]: assuming the ground state vortex lattice is pinned by the underlying network, the motion of the lattice is via a thermally activated domain nucleation process. The energy cost for creating a domain scales as $\sigma L$, where $\sigma$ is the wall energy per unit length and $L$ the length of the domain wall; the energy gain from transport current $I$ scales as $IL^2$. Thus the energy barrier for growing an unbounded domain scales as $I^{-1}$ and this process gives an exponential form $V \sim I \exp(-I_T/I)$ for $I$-$V$ curves. As discussed earlier, due to the symmetry of the ground states for non-integer fillings, both vortex pairs and domain walls are possible excitations. Note that the energy cost to creat free vortices from bound pairs diverges logarithmically with vanishing current, while the cost to grow an unbounded domain diverges algebraically. Therefore, if the two types of excitations are independent the low-temperature $I$-$V$ curves should be power-laws. The observed exponential $I$-$V$ curves are then either due to the coupling between the two types of excitations, or a result of a vanishing wall energy $\sigma \to 0$ close to the transition. For $f = 1/2$, we note that an early experiment [17] on Josephson arrays reported power-law $I$-$V$ curves at low temperatures, and a recent experiment [18] on Al wire networks showed $I$-$V$ curves that appear to be exponential. The origin of this difference between the two systems is not clear.

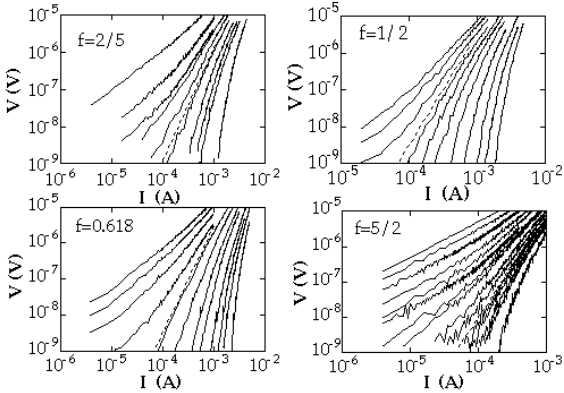

FIG. 3. Temperature dependent current-voltage characteristics at various fillings $f$ =2/5, 1/2, 0.618, and 5/2. The $I$-$V$ curves correspond to the following temperature ranges: for $f = 2/5$, $T$ =8.900 K (top $I$-$V$) to 8.850 K (bottom $I$-$V$); for $f = 1/2$, $T$ =8.880 K to 8.847 K; for $f = 0.618$, $T = 8.890$ K to 8.845 K; for $f = 5/2$, $T = 8.905$ K to 8.855 K. The dashed lines are hand-drawn to indicate the $V \sim I^3$ power-law.

The exponential form of $I$-$V$ characteristics suggests superconducting phase coherence at low temperatures in the wire network. For $f = 1/2$ and $2/5$, this may be understood as the long-range ordering in the pinned ground state vortex lattice. For higher-order commensurate or incommensurate fillings, e.g. $f = 0.618$, the existence of an ordered ground state could be an issue [6]. For adsorbate systems [7] a strong pinning potential can force an incommensurate lattice into registry with the substrate potential. The ground state is a pinned commensurate lattice [7]. However, in our system the vortex density is fixed by an external field and the screening is very weak in the temperature range of the experiment. The ground state of the system may be either commensurate domains separated by domain walls, or ordered quasiperiodically. The exponential $I$-$V$ curves found in this work and in a previous one [19] suggest a true superconducting state at low temperature for $f = 0.618$. This would imply that the vortex lattice is pinned at low temperature, either by the network or by the network irregularities which undoubtedly exist in our sample (a worst-case estimate from the SEM photo in Fig. 1 gives $\sim 10\%$ in areal fluctuations among the cells ) and that of the previous work [19].

The $I$-$V$ data in Fig. 3 suggest phase transitions from the high-temperature resistive state to low-temperature superconducting state. If the superconducting transitions are continuous, one would expect to see scaling behavior in $I$-$V$ characteristics [20,21]. In the following we attempt a scaling analysis of the $I$-$V$ data in Fig. 3. Approaching the transition, the phase correlation length $\xi$ (not to be confused with the bulk superconducting coherence length) in the network diverges as $\xi \sim |T - T_c|^{-\nu}$. Dimensionality arguments [21] suggest a scaling relation for the $I$-$V$ (or $j$-$E$) curves, in 2D, as $(E/j)|T - T_c|^{-z\nu} = G_\pm(j|T - T_c|^{-\nu})$, where $z$ is the dynamical critical exponent and $G_\pm$ the scaling functions above and below $T_c$. At $T = T_c$ and in 2D, $V \sim I^{z+1}$; thus we can determine $z$ directly from the $I$-$V$ curves in Fig. 3. For all the cases which we studied carefully, $z = 2$; this includes $f = 2/5$, $1/2$, $0.618$, and $5/2$. In the scaling analyses, we fix the parameter $z = 2$, as determined independently from the power law, and keep $\nu$ adjustable until a good collapse of data is achieved. The scaled data are shown in Fig. 4. The four sets of $I$-$V$ curves in Fig. 3 are found to show good scaling behavior, which strongly suggests a finite-temperature continuous superconducting transition for all these fillings. This is surprising because one would expect very different types of transitions for these different fillings.

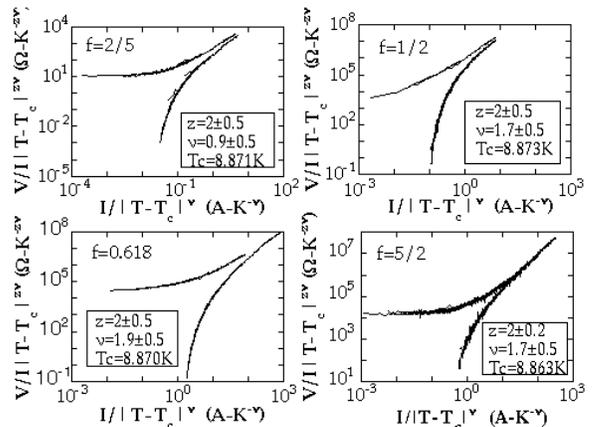

FIG. 4. Scaling plots for $f$ =1/2, 2/5, 0.618, and 5/2. The insets showing the values of $z$, $\nu$ and $T_c$ used to scale the data (see text).

At $f = 1/2$, one may have a continuous transition [9–11]. But for $f = 2/5$, one expects [22] a strongly first-order melting transition. At $f = 0.618$, a glass-like freezing into a metastable state may occur at a finite temperature, but no finite-temperature phase transition, thus no scaling behavior, is expected [6], in contrast to the interpretation in a previous work [19]. As discussed earlier, a true superconducting phase at low temperatures for $f = 0.618$ may be a result of the network irregularities. The fact that there is no pronounced dip feature at $f = 2/5$ on the $R$ vs. $f$ curves (Fig. 2) suggests that the $f = 2/5$ ground state may also be affected by such disorder. Since the first-order transitions can be driven continuous by quenched disorder [23,24], especially in 2D [23], it is possible that the continuous transitions for $f = 2/5$ [25] and 0.618 are due to the areal disorder in the network. For $f = 1/2$, we note that the critical exponent $\nu \approx 1.7$ is different from the 2D Ising value ($\nu = 1$) and from the nonuniversal value ($\nu \approx 0.8$) of a coupled $XY$-Ising model [11]. In random-bond 2D 8-state Potts model, the transition was found [24] to be driven by disorder to the Ising universality class. If weak disorder plays a role in driving a first-order transition continuous in our network, as conjectured above, an interesting question is whether the transitions at $f = 2/5$, $1/2$, and 0.618 are in the same universality class. Unfortunately, due to a systematic error ($\sim 4$ mK) in locating $T_c$, the error bar for exponent $\nu$ can be 0.3 to 0.5. Thus it is not clear whether the variation in $\nu$ for different fillings represents a real change in their universality classes, or they are in fact in a single universality class of either $f = 1/2$ or one controlled by disorder. In any case, the issue of disorder-induced continuous transition [23,24] in a 2D network deserves more careful studies and is currently under investigation.

Another remarkable feature of the data is the $V \sim I^3$ power-law for $f = 2/5$, $1/2$, and 0.618 at the transition, which is typical of a KT transition, but which occurs while the $I$-$V$ curves clearly suggest domain-wall excitations just below the transition. As mentioned earlier, the domain walls have the effect of screening the vortex-antivortex interactions, it is possible that an Ising-like transition at which the domain wall energy goes to zero may trigger a KT transition [10]. It is not clear, however, why the universal power of 3 which is related [26] to the universal jump in the helicity modulus of a KT transition should appear at such a transition.

We wish to thank S.N. Coppersmith, G. Grest, J.M. Kosterlitz, and S. Teitel for helpful discussions, A.M. Goldman and F. Yu for discussions of their work and for permission to quote their unpublished $f = 1/2$ data, and C. Denniston for informing us of his unpublished simulation results.